\documentclass[12pt]{article}
\usepackage{a4wide}
\usepackage{amsmath,amsthm,amsfonts,bbm}
\usepackage{graphicx,epsfig}
\usepackage{tikz}

\newcommand{\url}[1]{{\tt \small #1}}

\newcommand{\Ex}[2]{\mathbb{E}_{#1}\!\left[\,#2\,\right]}
\newcommand{\Qx}[1]{\mathbb{Q}\!\left\{\,#1\,\right\}}
\newcommand{\ind}[1]{\mathbbm{1}_{\{#1\}}}
\newcommand{\eqdef}{\mathrel{\mathop:}=}

\newcommand{\rec}{\mbox{R{\tiny EC}}}

\newcommand{\lgd}{\mbox{L{\tiny GD}}}

\newcommand{\npv}{\mbox{N{\tiny PV}}}

\newcommand{\cds}{\mbox{CDS}}
\newcommand{\brcva}{\mathrm{CVA\text{-}BR}}

\newcommand{\argmin}[1]{\underset{{#1}}{\operatorname{arg\,min}}\,}

\newtheorem{theorem}{Theorem}[section]
\newtheorem{definition}{Definition}[section]
\newtheorem{proposition}[theorem]{Proposition}
\newtheorem{remark}[theorem]{Remark}

\numberwithin{equation}{section}

\title{
\bf{\Large Bilateral counterparty risk valuation\\ for interest-rate products:\\ impact of volatilities and correlations.}}
\author{Damiano Brigo\thanks{Department of Mathematics, Imperial College, and
Fitch Solutions, 101 Finsbury Pavement, EC2A 1RS London, UK.
E-mail: {\tt d.brigo@imperial.ac.uk}}, \
 Andrea Pallavicini\thanks{Banca Leonardo, Milano, Italy,
{\tt andrea.pallavicini@bancaleonardo.com}}, \
Vasileios Papatheodorou\thanks{
Fitch Solutions, London, UK, E-mail: {\tt vasileios.papatheodorou@fitchsolutions.com}}
}

\date{\small First Version: Nov 17, 2009. This Version: \today
}

\begin{document}

\maketitle

\pagestyle{myheadings}
\markboth{}{{\footnotesize  D. Brigo, A. Pallavicini and V. Papatheodorou. Bilateral CVA for Interest Rate products}}

\begin{abstract}
The purpose of this paper is introducing rigorous methods and formulas for bilateral counterparty risk credit valuation adjustments (CVA's) on interest-rate portfolios. In doing so, we summarize the general arbitrage-free valuation framework for counterparty risk adjustments in presence of bilateral default risk, as developed more in detail in Brigo and Capponi (2008), including the default of the investor. We illustrate the symmetry in the valuation and show that the adjustment involves a long position in a put option plus a short position in a call option, both with zero strike and written on the residual net present value of the contract at the relevant default times. We allow for correlation between the default times of the investor and counterparty, and for correlation of each with the underlying risk factor, namely interest rates. We also analyze the often neglected impact of credit spread volatility. We include Netting in our examples, although other agreements such as Margining and Collateral are left for future work.
\end{abstract}

{\bf{AMS Classification Codes}}: 60H10, 60J60, 60J75, 62H20, 91B70 \\ \indent {\bf{JEL Classification Codes}}: C15, C63, C65, G12, G13 \\

\medskip

{\bf Keywords:} Counterparty Risk, Arbitrage-Free Credit Valuation Adjustment, Interest Rate Swaps, Interest Rate Derivatives, Credit Valuation Adjustment, Bilateral Risk, Credit Spread Volatility, Default Correlation, Stochastic Intensity, Short Rate Models, Copula Functions, Wrong Way Risk.

\newpage
{\small \tableofcontents}
\newpage

\section{Introduction}

This paper deals with pricing of bilateral counterparty risk credit valuation adjustments (CVA's) on interest-rate portfolios. The structure of the paper is similar to the earlier work of Brigo and Capponi (2008). However, while the focus in Brigo and Capponi (2008) is on Credit Default Swaps (CDS), here it is on interest-rate products, generalizing to the bilateral case the earlier works on unilateral CVA for rates products done by Sorensen and Bollier (1994), Brigo and Masetti (2005) and Brigo and Pallavicini (2007). Indeed, previous research on accurate arbitrage-free valuation of unilateral CVA with dynamical models on commodities (Brigo and Bakkar (2009)), on rates (Brigo and Pallavicini (2007)) and on credit (Brigo and Chourdakis (2008)) assumed the party computing the valuation adjustment to be default-free. We present here the general arbitrage-free valuation framework for counterparty risk adjustments in presence of bilateral default risk, as introduced in Brigo and Capponi (2008),  including default of the investor. We illustrate the symmetry in the valuation and show that the adjustment involves a long position in a put option plus a short position in a call option, both with zero strike and written on the residual net value of the contract at the relevant default times. We allow for correlation between the default times of the investor, counterparty and underlying portfolio risk factors. We use arbitrage-free stochastic dynamical  models.

We then specialize our analysis to Interest rate payouts as underlying portfolio. In comparing with the CDS case as underlying, an important point is that most credit models in the industry, especially when applied to Collateralized Debt Obligations or $k$-th to default baskets, model default correlation but ignore credit-spread volatility. Credit spreads are typically assumed to be deterministic and a copula is postulated on the exponential triggers of the default times to model default correlation. This is the opposite of what used to happen with counterparty risk for interest-rate underlyings, for example in Sorensen and Bollier (1994) or Brigo and Masetti (in Pykhtin (2005)), where correlation was ignored and volatility was modeled instead. Brigo and Chourdakis (2008) rectify this in the CDS context, but only deal with unilateral and asymmetric counterparty risk. Brigo and Capponi (2008) then generalize this approach for CDS, including credit spread volatility besides default correlation into the bilateral case.

In interest-rate products, previous literature dealing with both underlying volatility and underlying/counterparty correlation is in Brigo and Pallavicini  (2007), who address both plain vanilla interest-rate swaps and exotics under unilateral counterparty risk. In that work, a stochastic intensity model along the lines of Brigo and Alfonsi (2005) and Brigo and El-Bachir (2009) is assumed, and this model is correlated with the multi-factor short rate process driving the interest-rate dynamics. Netting is also examined in some basic examples. The present paper aims at generalizing this approach to the bilateral case. In such a case one needs to model the following correlations, or better dependencies:
\begin{itemize}
\item Dependence between default of the counterparty and default of the investor;
\item Correlation between the underlying (interest rates) and the counterparty credit spread;
\item Correlation between the underlying (interest rates) and the investor credit spread;
\item Besides default correlation between the counterparty and the investor, we might wish to model also credit spread correlation.
\end{itemize}
We will model all such dependencies except the last one, since default correlation is dominant over spread correlation.
Also, a feature that is usually ignored is credit spread volatility for the investor and the counterparty, in that credit spreads are usually taken as deterministic. We improve this by assuming stochastic spreads for both investor and counterparty.

Summarizing, to avoid confusion, given the plurality of papers we produced in recent years, we can put this paper's contribution into context with respect to our analogous earlier versions for unilateral counterparty risk or for other asset classes through Table~\ref{table:contextCRmodels}, which is clearly meant only as a partial orientation for our work and not for the overall much broader literature.

\begin{table}
\begin{center}
{\small
\begin{tabular}{|c||c|c|c|}
\hline
 CVA & \multicolumn{ 2}{|c|}{One-sided} &  Bilateral \\\hline
Modeling & Volatility & \multicolumn{ 2}{|c|}{Volatility and Correlation} \\\hline\hline
IR Swaps &  Brigo Masetti (2005)         & Brigo Pallavicini (2007) & This paper \\
{\scriptsize with Netting} &  & & \\\hline
IR Exotics &            & Brigo Pallavicini (2007) & This paper \\\hline
 Oil Swaps &            & Brigo Bakkar (2009) &            \\\hline
       CDS &            & Brigo Chourdakis (2008) & Brigo Capponi (2008) \\ \hline
Equity TRS &            & Brigo Tarenghi (2004) &            \\
           &            & Brigo Masetti (2005) &            \\ \hline
\end{tabular}}
\end{center}
\caption{Part of earlier analogous literature on CVA valuation with respect to inclusion of underlying asset volatilities and/or correlation between underlying asset and counterparties, along with bilateral features. TRS stands for Total Return Swap}\label{table:contextCRmodels}
\end{table}

We specify that we do not consider specific collateral clauses or guarantees in the present work, although we deal with some stylized cases of netting. We assume we are dealing with counterparty risk for an over the counter interest-rate portfolio transaction where there is no periodic margining or collateral posting. Past works where  netting has been addressed in the interest-rate context are Brigo and Pallavicini (2007) and Brigo and Masetti (2005). The impact of credit triggers for the counterparty on CVA are analyzed in Yi (2009). Assefa et al (2009) analyze the modeling of collateralization and margining in CVA calculations.

Finally, given the theoretical equivalence of the credit valuation adjustment with a contingent CDS, we are also proposing a methodology for valuation of contingent CDS on interest rates. See Brigo and Pallavicini (2007) for more details on contingent CDS.

\medskip

The paper is structured as follows:

Sections \ref{sec:generalcounterformula} summarizes the bilateral counterparty risk valuation formula from Brigo and Capponi (2008), establishing also the appropriate notation. A discussion on the specific features of bilateral risk and of some seemingly paradoxical aspects of the same, also in connection with real banking reports of 2009, is presented.

Section \ref{sec:IRSwap_application} details the application of the methodology to Interest-Rate Swaps. A two-factor Gaussian interest-rate model is proposed to deal with the option features of the bilateral counterparty risk adjustment. The model is calibrated to the zero curve and to swaptions. Then shifted square root diffusion credit spread models with possible jumps for both the counterparty and the investor are introduced. The defaults of the counterparty and of the investor are linked by a Gaussian copula. The correlation structures originating dependence between interest rates and defaults are explained in detail, and finally the numerical Monte Carlo techniques used to value the adjustment are illustrated.

Section \ref{sec:casestudyrates} presents a case study based on three possible interest-rate swaps portfolios, some embedding netting clauses. We analyze the impact of credit spread levels and volatilities, of correlations between the underlying interest rates and defaults, and of dependence between default of the counterparty and of the investor. Section \ref{sec:conclbilateralrates} concludes the paper.


\section{Arbitrage-free valuation of bilateral counterparty risk}
\label{sec:generalcounterformula}

The bilateral counterparty risk is mentioned in the Basel II documentation.

\begin{remark} {\bf (Bilateral Counterparty Risk in Basel II, Annex IV, 2/A)}
``Unlike a firm's exposure to credit risk through a loan, where the exposure to credit risk is unilateral and only the lending bank faces the risk of loss, the counterparty credit risk creates a bilateral risk of loss: the market value of the transaction can be positive or negative to either counterparty to the transaction."
\end{remark}

Basel II is more concerned with Risk Measurement than pricing. For an analysis of counterparty risk in the risk-measurement space we refer for example to De Prisco and Rosen (2005), who consider modeling of stochastic credit exposures for derivatives portfolios.  However, also in the valuation space, bilateral features are quite relevant and often can be responsible for seemingly paradoxical statements\footnote{We are grateful to Dan Rosen for first signaling this issue to us during a conference in June 2009}, as pointed out in Brigo and Capponi (2008). For example, Citigroup in its press release on the first quarter revenues of 2009 reported a {\em positive} mark to market due to its {\em worsened} credit quality: ``Revenues also included [...] a net 2.5\$ billion positive CVA on derivative positions, excluding monolines, mainly due to the widening of Citi's CDS spreads". In this paper we explain precisely how such a situation may origin.

We refer to the two names involved in the transaction and subject to default risk as
\begin{eqnarray}
\nonumber \text{investor } & \rightarrow & \text{name ``I''} \\
\nonumber \text{counterparty } & \rightarrow & \text{name ``C''}
\label{eq:three_names}
\end{eqnarray}

In general, we will address valuation as seen from the point of view of the investor ``I", so that cash flows received by ``I" will be positive whereas cash flows paid by ``I" (and received by ``C") will be negative\footnote{Here, we follow Brigo and Capponi (2008), although in that paper the investor name ``I" is called ``0" and the counterparty name ``C" is called ``2".}. 

We denote by $\tau_I$ and $\tau_C$ respectively the default times of the investor and counterparty. We place ourselves in a probability space $(\Omega,\mathcal{G},\mathcal{G}_t,\mathbb{Q})$. The filtration $\mathcal{G}_t$ models the flow of information of the whole market, including credit and $\mathbb{Q}$ is the risk neutral measure. This space is endowed also with a right-continuous and complete sub-filtration $\mathcal{F}_t$ representing all the
observable market quantities but the default events, thus $\mathcal{F}_t\subseteq\mathcal{G}_t\eqdef\mathcal{F}_t\vee\mathcal{H}_t$. Here, $\mathcal{H}_t=\sigma(\{\tau_I\leq u \} \vee \{\tau_C\leq u\} :u\leq t)$ is the right-continuous filtration generated by the default events,
either of the investor or of his counterparty.

Let us call $T$ the final maturity of the payoff which we need to evaluate and let us define the stopping time
\begin{equation}
\tau = \min\{\tau_I, \tau_C\}
\label{eq:stopping_time}
\end{equation}
If $\tau>T$, there is neither default of the investor, nor of his counterparty during the life of the contract and they both can fulfill the agreements of the contract. On the contrary, if $\tau\leq T$ then either the investor or his counterparty (or both) default. At $\tau$, the Net Present Value (NPV) of the residual payoff until maturity is computed. We then distinguish two cases:

\begin{itemize}
\item $\tau = \tau_C$. If the NPV is negative (respectively positive) for the investor (defaulted counterparty), it is completely paid (received) by the investor (defaulted counterparty) itself. If the NPV is positive (negative) for the investor (counterparty), only a recovery fraction $\rec_C$ of the
NPV is exchanged.
\item $\tau = \tau_I$. If the NPV is positive (respectively negative)  for the defaulted investor (counterparty), it is completely received (paid) by the defaulted investor (counterparty) itself. If the NPV is negative (positive) for the defaulted investor (counterparty), only a recovery fraction $\rec_I$ of the NPV is exchanged.
\end{itemize}

Let us define the following (mutually exclusive and exhaustive) events ordering the default times
\begin{eqnarray}
\label{eq:event_set}
\nonumber A &=& \{\tau_I \le \tau_C \le T\} \quad\quad E = \{T \le \tau_I \le \tau_C \} \\
\nonumber B &=& \{\tau_I \le T \le \tau_C\} \quad\quad F = \{T \le \tau_C \le \tau_I \} \\
\nonumber C &=& \{\tau_C \le \tau_I \le T \} \\
D &=& \{\tau_C \le T \le \tau_I \}
\end{eqnarray}

Let us call $\Pi^D(t,T)$ the discounted payoff of a generic defaultable claim at $t$ and $\Pi(t,T)$ the discounted payoff for an equivalent claim with a default-free counterparty. We then have the following Proposition, proven in Brigo and Capponi (2008)
\begin{proposition}
\label{prop_main}
{\bf (General bilateral counterparty risk pricing formula)}
At valuation time $t$, and conditional on the event $\{ \tau > t\}$, the price of the payoff under bilateral counterparty risk is
{\emph{
\begin{eqnarray}
\label{generalprice}
\Ex{t}{\Pi^D(t,T)} & = & \Ex{t}{\Pi(t,T)} \nonumber \\
& & + \,\Ex{t}{ \lgd_I \ind{A \cup B} D(t,\tau_I) \left(-\npv(\tau_I)\right)^+ } \nonumber \\
& & - \,\Ex{t}{ \lgd_C \ind{C \cup D} D(t,\tau_C) \left( \npv(\tau_C)\right)^+ }
\end{eqnarray}}}
where {\emph{$\lgd_i \eqdef 1 - \rec_i$}} is the \emph{Loss Given Default} and {\emph{$\rec_i$}} is the recovery fraction, with $i\in\{I,C\}$. It is clear that the value of a defaultable claim is the value of the corresponding default-free claim plus a long position in a put option (with zero strike) on the residual NPV giving nonzero contribution only in scenarios where the investor is the earliest to default (and does so before final maturity) plus a short position in a call option (with zero strike) on the residual NPV giving non-zero contribution in scenarios where the counterparty is the earliest to default (and does so before final maturity).
\end{proposition}

\begin{definition}{\bf (Bilateral CVA, DVA, CVA)}
The adjustment is called bilateral counterparty-risk credit-valuation adjustment and it may be either negative or positive depending on whether the counterparty is more or less likely to default than the investor and on the volatilities and correlation. From the investor point of view, we define $\brcva$ the adjustment to be {\em added} to the default free price to account for counterparty risk.

When looking at the adjustment from the point of view of the investor ``I", in the right hand side of~(\ref{generalprice}) the second term and the third term being subtracted from the second one are called respectively (unilateral) Debit Valuation Adjustment (DVA) and (unilateral) Credit Valuation Adjustment (CVA), so that the mathematical expression for the bilateral adjustment\footnote{In Brigo and Capponi (2008) tables report the opposite quantity, $-\brcva$=:BR-CVA=CVA-DVA.} is given by
\begin{eqnarray}
\label{eq:br-cva}
\brcva(t,T) &=& \mbox{DVA}(t,T) - \mbox{CVA}(t,T) , \\
  \mbox{DVA}(t,T)&=& \,\Ex{t}{ \lgd_I \ind{A \cup B} \cdot D(t,\tau_I) \cdot \left( -\npv(\tau_I)\right)^+ } \nonumber \\
\mbox{CVA}(t,T) &=& \Ex{t}{ \lgd_C \ind{C \cup D} \cdot D(t,\tau_C) \cdot \left(\npv(\tau_C)\right)^+ }
\end{eqnarray}
where the right hand side in Equation \eqref{eq:br-cva} depends on $T$ through the events $A,B,C,D$ and $\lgd_i$, with $i\in\{I,C\}$, is a shorthand notation to denote the dependence on the loss given defaults of each name.
\end{definition}

Notice that in the paper we assume the recovery fractions (and hence LGD's) to be deterministic.

\begin{remark}
{\bf (Symmetry vs Asymmetry).}
With respect to earlier results on counterparty risk valuation, Equation \eqref{eq:br-cva} has the great advantage of being symmetric. This is to say that if ``C" were to compute counterparty risk of her position towards ``I", i.e. the term to be added to the default free price to include counterparty risk, she would find exactly $-\brcva(t,T)$. However, if each party computed the adjustment to be added by assuming itself to be default-free and considering only the default of the other party, then the adjustment calculated by ``I'' would be
{\emph{
\[
-\,\Ex{t}{ \lgd_C \ind{\tau_C < T} \cdot D(t,\tau_C) \cdot \left(\npv(\tau_C)\right)^+ }
\]}}
whereas the adjustment calculated by ``C'' would be
{\emph{
\[
-\,\Ex{t}{ \lgd_I \ind{\tau_I < T} \cdot D(t,\tau_I) \cdot \left(-\npv(\tau_I)\right)^+ }
\]}}
and they would not be one the opposite of the other. This means that only in the first case the two parties agree on the value of the counterparty risk adjustment to be added to the default-free price.
\end{remark}

\begin{remark}
{\bf (Change in sign).}
Earlier results on asymmetric counterparty risk valuation, concerned with a default-free investor, would find an adjustment to be added that is always negative. However, in our symmetric case even if the initial adjustment is negative due to CVA$(t,T)>$DVA$(t,T)$, i.e.
{\emph{
\[
\Ex{t}{ \lgd_C \ind{C \cup D} \cdot D(t,\tau_C) \cdot \left(\npv(\tau_C)\right)^+ } >
   \Ex{t}{ \lgd_I \ind{A \cup B} \cdot D(t,\tau_I) \cdot \left(-\npv(\tau_I)\right)^+ }
\]}}
the situation may change in time, to the point that the two terms may cancel or that the adjustment may change sign as the credit quality of ``I" deteriorates and that of ``C" improves, so that the inequality changes direction.
\end{remark}

\begin{remark}
{\bf (Worsening of credit quality and positive mark to market).}
If the Investor marks to market her position at a later time using Equation \eqref{generalprice}, we can see that the term in $\lgd_I$ increases, ceteris paribus, if the credit quality of ``I" worsens. Indeed, if we for example increase the credit spreads of the investor, now $\tau_I < \tau_C$ will happen more often, giving more weight to the term in $\lgd_I$. This is at the basis of statements like the above one of Citigroup.
\end{remark}

\section{Application to Interest Rate Products} \label{sec:IRSwap_application}

In this section we consider a model that is stochastic both in the interest rates (underlying market) and in the default intensity (counterparty). Joint stochasticity is needed to introduce correlation. The interest-rate sector is modeled according to a short-rate Gaussian shifted two-factor process (hereafter G2++), while each of the two default-intensity sectors is modeled according to a square-root process with exponential jumps (hereafter JCIR++). Details for both models can be found, for example, on Brigo and Mercurio (2006). The two models are coupled by correlating
their Brownian shocks.

\subsection{Interest rate model: G2++}\label{subsec:interestrates}

For interest rates, we assume that the dynamics of the instantaneous short-rate process under the risk-neutral measure is given by
\begin{equation}
\label{ch4:hw2srdyn}
r(t) = x(t) + z(t) + \varphi(t;\alpha) \;,\quad r(0)=r_0,
\end{equation}
where $\alpha$ is a set of parameters and the processes $x$ and $z$ are ${\cal F}_t$ adapted and satisfy
\begin{equation}
\label{ch4:hw2xydyn}
\begin{split}
dx(t) & = -a x(t) dt + \sigma dZ_1(t) \;,\quad x(0)=0,\\
dz(t) & = -b z(t) dt + \eta dZ_2(t) \;,\quad z(0)=0,
\end{split}
\end{equation}
where $(Z_1,Z_2)$ is a two-dimensional Brownian motion with instantaneous correlation $\rho_{12}$ as from
\[
d \langle Z_1, Z_2\rangle_t = \rho_{12} dt,
\]
where $r_0$, $a$, $b$, $\sigma$, $\eta$ are positive constants, and where $-1 \le \rho_{12} \le 1$. These are the parameters entering $\varphi$, in that $\alpha = [r_0, a, b, \sigma, \eta, \rho_{12}]$. The function $\varphi(\cdot;\alpha)$ is deterministic and well defined in the time interval $[0,T^*]$, with $T^*$ a given time horizon, typically 10, 30 or 50 (years). In particular, $\varphi(0;\alpha)=r_0$. This function can be set
to a value automatically calibrating the initial zero coupon curve observed in the market.

We calibrate the interest-rate model parameters to the ATM swaption volatilities quoted by the market on May 26, 2009. Market data are listed in Appendix~\ref{app:rates}, while more details on the methodology can be found on Brigo and Pallavicini (2007). Below, we report the calibrated model parameters and absolute calibration errors in basis points (expiries on the left axis, tenors on the right axis).

\begin{center}
\begin{minipage}{0.48\textwidth}
\centering
{\small
\begin{tabular}{|cc|}\hline
$a$ & 0.0002 \\
$b$ & 7.6630 \\
$\sigma$ & 0.0080 \\
$\eta$ & 0.0182 \\
$\rho_{12}$ & 0.9734 \\\hline
\end{tabular}}
\end{minipage}
\begin{minipage}{0.48\textwidth}
\centering
\includegraphics[width=\textwidth]{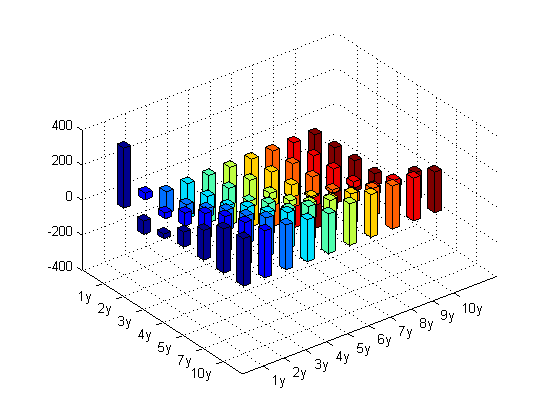}
\end{minipage}
\end{center}

The G2++ model links the dependence on tenors of swaption volatilities to the form of initial yield curve. Before the crisis period such constraint of the G2++ model seems not so relevant, but the situation changes from spring 2008, when the yield curve steepened in conjunction with a movement in the market volatility surface which could not be reproduced by the model. Yet, versions of the model with time-dependent volatilities can calibrate ATM swaption volatilities in a satisfactory way. For instance, if we introduce a time grid $t_0=0,t_1,\ldots,t_m$, we can consider the following time-dependent volatilities.
\[
\sigma(t) \eqdef {\bar\sigma} f(\ell(t))
\;,\quad
\eta(t) \eqdef {\bar\eta} f(\ell(t))
\]
where the $\ell(t) \eqdef \max\{t^*\in\{t_0,\ldots,t_m\} : t^*\le t \}$ function selects the left extremum of each interval and
\[
f(t) \eqdef 1 - e^{-\beta_1t} + \beta_0e^{-\beta_2t}
\]
Notice that in this way we do not alter the analytical tractability of the G2++ model, since all integrals involving model piece-wise-constant parameters can be performed as finite summations.

However, in this presentation on counterparty risk we consider the simpler constant parameter version of the G2++ model. Nonetheless, we report below model parameters and absolute calibration errors in basis points for the time-dependent version of the G2++ model (expiries on the left axis, tenors on the right axis).
\begin{center}
\begin{minipage}{0.48\textwidth}
\centering
{\small
\begin{tabular}{|cc|}\hline
$a$ & 0.0001 \\
$b$ & 1.9478 \\
$\sigma$ & 0.0062 \\
$\eta$ & 0.0299 \\
$\rho_{12}$ & -0.7661 \\\hline
$\beta_0$ & 1.6241 \\
$\beta_1$ & 9.0793 \\
$\beta_2$ & 1.7074 \\\hline
\end{tabular}}
\end{minipage}
\begin{minipage}{0.48\textwidth}
\centering
\includegraphics[width=\textwidth]{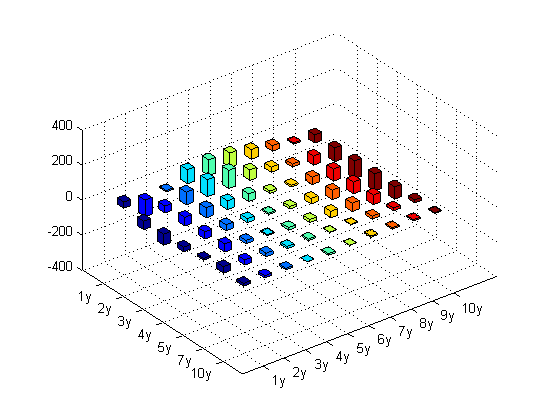}
\end{minipage}
\end{center}

\subsection{Counterparty and Investor Credit Spread models}\label{subsec:creditspreads}

For the stochastic intensity models we set
\begin{equation}
\label{extended_i}
\lambda^{i}_t = y^{i}_t + \psi^{i}(t;\beta^i) \;,\quad i\in \{I,C\}
\end{equation}
where whenever we omit the upper index we refer to quantities for both indices. The function $\psi$ is a deterministic function, depending on the parameter vector $\beta$ (which includes $y_0$), that is integrable on closed intervals. The initial condition $y_0$ is one more parameter at our disposal. We are free to select its value as long as
\[
\psi(0;\beta) = \lambda_0 - y_0 \ .
\]
We take $y$ to be a Cox Ingersoll Ross process with exponentially distributed positive jumps (see for example Brigo and Mercurio (2006)):
\[
dy^i_t = \kappa^i(\mu^i-y^i_t)dt + \nu^i \sqrt{y^i_t}dZ^i_3(t) + dJ^i_t(\zeta^i_1,\zeta^i_2) \;,\quad i \in \{I,C\}
\]
where the parameter vector is $\beta^i \eqdef (\kappa^i,\mu^i,\nu^i,y^i_0,\zeta^i_1,\zeta^i_2)$ and each parameter is a positive deterministic constant. As usual, $Z_3^i$ is a standard Brownian motion process under the risk neutral measure, while the jump part $J_t(\zeta_1,\zeta_2)$ is defined as
\[
J^i_t(\zeta^i_1,\zeta^i_2) \eqdef \sum_{k=1}^{M^i_t(\zeta^i_1)}X^i_k(\zeta_2^i) \;,\quad i \in \{I,C\}
\]
where $M^i$ is a time-homogeneous Poisson process (independent of $Z$) with intensity $\zeta^i_1$, the $X^i$s being exponentially distributed with positive finite mean $\zeta^i_2$ independent of $M$ (and $Z$). The two processes $y^I$ and $y^C$ are assumed to be independent, so that $Z_3^I$ is independent of $Z_3^C$ and $J^I$ is independent of $J^C$ (in particular, $M^I$ is independent of $M^C$ and $X^I$'s of $X^C$'s) . This is assumed to simplify the parametrization of the model and focus on default correlation rather than spread correlation, but the assumption can be removed if one is willing to complicate the parametrization of the model.

We define the integrated quantities
\[
\Lambda(t) \eqdef \int_0^t\lambda_s ds
\;,\quad
Y(t) \eqdef \int_0^t y_s ds
\;,\quad
\Psi(t,\beta) \eqdef \int_0^t \psi(s,\beta)ds \ .
\]

In our Cox process setting the default times are modeled as
\[
\tau_i = (\Lambda^i)^{-1}(\xi_i) \;,\quad i \in \{I,C\}
\]
with $\xi$'s each exponential unit-mean and independent of interest rates. The two $\xi$ are assumed to be connected via a bivariate Gaussian copula function with correlation parameter $\rho_G$. This is a default correlation, and the two default times are connected via default correlation, even if their spreads are independent. In fact, in general high default correlation creates more dependence between the default times than a high correlation in their spreads.

From this setup it follows that when we assume the default intensity $\lambda$, and the cumulated intensity
$\Lambda$, to be independent of the short rate $r$ and of interest rates in general, also the default times $\tau_i$ will be independent of interest-rate related quantities $r, D(s,t),...$. In this case valuation of (running) CDS on reference entities ``I" or ``C" becomes model independent, leading to
{\small
\begin{eqnarray}
\label{ch:credit:modindcdstot}
\cds^i_{a,b}(0,S^i) &=&
\lgd \left[ \int_{T_a}^{T_b} P(0,t) \,d_t \Qx{\tau_i \ge t} \right] \\ \nonumber
  & & + S^i \left[ -\int_{T_a}^{T_b} P(0,t) (t-T_{\gamma(t)-1}) \,d_t\Qx{\tau_i \ge t}
      + \sum_{j=a+1}^b \alpha_j P(0,T_j) \,\Qx{\tau_i \ge T_j} \right]
\end{eqnarray}}
%
where in general $T_{\gamma(t)}$ is is the first $T_j$ following $t$ and $P(t,T)=\Ex{t}{D(t,T)}$ is the zero coupon bond price at time $t$ for maturity $T$ consistent with the stochastic discount factors $D$. This formula is model independent, see for example the credit chapters in Brigo and Mercurio (2006) for the details, $S$ is the CDS spread in the premium leg, typically balancing the default leg at inception. For conversion of these running CDS into upfront ones, following the so called Big Bang protocol by ISDA, see for example Beumee, Brigo, Schiemert and Stoyle (2009). Since the survival probabilities in the JCIR++ model are given by
\begin{equation}\label{ch:credit:cdscalrho0}
\Qx{\tau>t}_{\rm \tiny model} = \Ex{0}{\exp\left(-\Lambda(t)\right)} = \Ex{0}{\exp \left(-\Psi(t,\beta)-Y(t) \right)}
\end{equation}
we just need to make sure
\begin{equation*}
\Ex{0}{\exp\left(-\Psi(t,\beta) - Y(t)\right)} = \Qx{\tau>t}_{\rm \tiny CDSmarket}
\end{equation*}
from which
\begin{equation}
\label{fittingPsicir}
\Psi(t,\beta) =
\ln\left(\frac{\Ex{0}{e^{-Y(t)}}}{\Qx{\tau>t}_{\rm \tiny CDSmarket}}\right) =
\ln\left(\frac{P_{\rm \tiny JCIR}(0,t, y_0; \beta)}{\Qx{\tau>t}_{\rm \tiny CDSmarket}}\right)
\end{equation}
where we choose the parameters $\beta$ in order to have a positive function $\psi$ (i.e. an increasing $\Psi$) and $P_{\rm \tiny JCIR}$ is the closed form expression for bond prices in the time-homogeneous JCIR model with initial condition $y_0$ and parameters $\beta$ (see for example Brigo and El-Bachir (2009), reported also in Brigo and Mercurio (2006)). Thus, if $\psi$ is selected according to this last formula, as we will assume from now on, the model is easily and automatically calibrated to the market survival probabilities (possibly stripped from CDS data).

This CDS calibration procedure assumes zero correlation between default and interest rates, so in principle when taking non-zero correlation we cannot adopt it. However, we have seen in Brigo and Alfonsi (2005) and further in Brigo and Mercurio (2006) that the impact of interest-rate / default correlation is typically small on CDSs, so that we may retain this calibration procedure even under non-zero correlation.

Once we have done this and calibrated CDS data through $\psi(\cdot,\beta)$, we are left with the parameters $\beta$,  which can be used to calibrate further products. However, this will be interesting when single name option data on the credit derivatives market will become more liquid. Currently the bid-ask spreads for single name CDS options are large and suggest to consider these quotes with caution, see Brigo (2005). At the moment we content ourselves of calibrating only CDS's for the credit part. To help specifying $\beta$ without further data we set some values of the
parameters implying possibly reasonable values for the implied volatility of hypothetical CDS options on the counterparty ``C" and investor ``I". Further, we always consider for the following numerical results that the jump part of the model is switched off. See Brigo and Pallavicini (2007) to size the impact of jumps.

We focus on two different sets of CDS quotes, that we name hereafter {\em Mid} and {\em High} risk settings. Then, we introduce a different set of model parameters for each CDS setting. In the following tables we show them along with the implied volatilities for CDSs starting at $t$ and maturing at $T$. The implied volatilities are calculated via a Jamshidian's decomposition as described in Brigo and Alfonsi (2005) or Brigo and Mercurio (2006).

The interest-rate curve is bootstrapped from the market on May 26, 2009 (see Appendix~\ref{app:rates}). Notice that the zero-curve is increasing in time. Further, we always consider that recovery rates are at $40\%$ level.

We consider two market settings for the credit quality and volatility of ``I" and ``C": a mid-risk setting and a high risk setting. The mid risk setting parameters are given in Table~\ref{tab:yparammid1}, and the associated CDS term structure and implied volatilities are reported in Appendix~\ref{app:credit}.
\begin{table}[htp]
\begin{center}
\begin{tabular}{|c|c|c|c|}\hline
  $\boldsymbol {y_0}$ & $\boldsymbol\kappa$ & $\boldsymbol\mu$ & $\boldsymbol\nu$ \\\hline
  0.01 & 0.80 & 0.02 & 0.20 \\\hline
\end{tabular}
\caption{Mid risk Credit spread parameters}\label{tab:yparammid1}
\end{center}
\end{table}

\noindent
The High risk market parameters are in Table~\ref{tab:yparamhigh1} and the associated CDS term structure and implied volatilities are reported in Appendix~\ref{app:credit}.
\begin{table}[htp]
\begin{center}
\begin{tabular}{|c|c|c|c|}\hline
  $\boldsymbol {y_0}$ & $\boldsymbol\kappa$ & $\boldsymbol\mu$ & $\boldsymbol\nu$ \\\hline
  0.03 & 0.50 & 0.05 & 0.50 \\\hline
\end{tabular}
\caption{High risk Credit spread parameters}\label{tab:yparamhigh1}
\end{center}
\end{table}

\subsection{Interest-rate / credit-spread correlations}

We take the short interest-rate factors $x$ and $z$ and the intensity process $y$ to be correlated, by assuming the driving Brownian motions $Z_I, Z_C$ and $Z_3$ to be instantaneously correlated according to
\[
d \langle Z_j ,Z^i_3 \rangle_t  = \rho_{j,i} dt \;,\quad j \in \{1,2\} \;,\; i \in \{I,C\}
\]

Notice that the instantaneous correlation between the resulting short-rate and the intensity, i.e. the instantaneous interest-rate / credit-spread correlation is
\[
\bar{\rho}_i \eqdef
\frac{d \langle r , \lambda^i \rangle_t}{\sqrt{d \langle r ,r \rangle_t\ d \langle \lambda^i , \lambda^i \rangle_t}} =
\frac{\sigma \rho_{1i}+\eta \rho_{2i}}
     {\sqrt{\sigma^2 + \eta^2 + 2 \sigma \eta \rho_{12}} \sqrt{1+\frac{2\zeta^i_1 \zeta^i_2}{(\nu^i)^2 y_t} }}
\;,\quad i \in \{I,C\} \ .
\]
%
This is a state dependent quantity due to presence of jumps. Without jumps, this simplifies to
\[
\bar{\rho}_i =
\frac{\sigma \rho_{1i}+\eta \rho_{2i}}{\sqrt{\sigma^2 + \eta^2 + 2 \sigma \eta \rho_{12}}} \ .
\]

In order to reduce the number of free parameters and to model in a more robust way the correlation structure of the model, in the following we always consider that
\[
\rho_{1i} = \rho_{2i}
\;,\quad i \in \{I,C\} \ .
\]
Further, we prefer to model default correlation by introducing a Gaussian copula on default times, rather than by correlating the default intensities, so that as explained above we take the two spread processes $y^I$ and $y^C$ to be independent.

\subsection{Monte Carlo techniques}

A Monte Carlo simulation is used to value all the payoffs.

The transition density for the G2++ model is known in closed form, while the JCIR++ model, even if we consider the jump part switched off, when correlated with G2++, requires a discretization scheme for the joint evolution. We find similar
convergence results both with the full truncation scheme introduced by Lord et al. (2006) and with the implied scheme by Brigo and Alfonsi (2005). In the following we adopt the former scheme.

Further, we bucket default times by assuming that the default events can occur only on a time grid $\{T_i:0 \leq i\le\ b\}$, with $T_0=t$ and $T_b=T$, by anticipating each default event to the last $T_i$ preceding it. In the following calculations we choose a weekly interval and we check a-posteriori that the time-grid spacing is small enough to have a stable value for the $\brcva$ price.

The calculation of the future time expectation, required by counterparty risk evaluation, is taken by approximating the
expectation at the actual (bucketed) default time $T_i$ with a finite series in the interest-rate model underlyings, $x$ and $z$, on a polynomial basis $\{\psi_j\}$ valued at the allowed default times within the interval $[t,T_i[$.
\[
\npv(T_i) \eqdef \Ex{T_i}{\Pi(T_i,T)} = \sum_{j=0}^\infty \alpha_{ij} \psi_j(x_{t:T_i},z_{t:T_i}) \simeq \sum_{j=0}^N \alpha_{ij} \psi_j(x_{t:T_i},z_{t:T_i})
\]
Notice that, if the payoff is not time-dependent, the functions $\psi$s need to be valued only at $T_i$.
The coefficients $\alpha_{ij}$ of the series expansion are calculated by means of a least-square regression, as usually done to price Bermudan options with the Least Squared Monte Carlo method.

Thus, the credit valuation adjustment is calculated as follows
\begin{eqnarray*}
\brcva(t,T)
&\simeq & - \, \lgd_C \sum_{i=0}^{b-1}
\Ex{t}{ \ind{\tau_I\geq\tau_C}\ind{T_i\leq\tau_C<T_{i+1}} D(t,T_i) \left(\Ex{T_i}{\Pi(T_i,T)}\right)^+} \\
&      & + \, \lgd_I \sum_{i=0}^{b-1} \Ex{t}{\ind{\tau_I\leq\tau_C}\ind{T_i\leq\tau_I<T_{i+1}} D(t,T_i) \left(-\,\Ex{T_i}{\Pi(T_i,T)}\right)^+}
\end{eqnarray*}
where the forward expectations are approximated as
\[
\Ex{T_i}{\Pi(T_i,T)} \simeq \sum_{j=0}^N \alpha_{ij} \psi_j(x_{t:T_i},z_{t:T_i})\\
\]
\[
\{\alpha_{ij}\} = \argmin{\alpha_{i0},\dots,\alpha_{iN}} \Ex{t}{\Big( \Pi(T_i,T) - \sum_{j=0}^N \alpha_{ij} \psi_j(x_{t:T_i},z_{t:T_i}) \Big)^2}
\]

In the following numerical examples we consider non-path-dependent payoffs, and we empirically find stable prices by using a polynomial basis up to the second degree in the function parameters, namely
\begin{gather*}
\psi_{0}(x,z) \eqdef 1 \;,\quad \psi_{1}(x,z) \eqdef x  \;,\quad \psi_{2}(x,z) \eqdef z \\
\psi_{3}(x,z) \eqdef x^2 \;,\quad \psi_{4}(x,z) \eqdef z^2  \;,\quad \psi_{5}(x,z) \eqdef x z
\end{gather*}
Notice also that, since the payoff evaluation depends on the projection coefficients which, in turn, depend on the simulated path, we are introducing a correlation between our Monte Carlo samples which, in principle, makes the standard deviation a biased estimator of the statistical error. However, in our experience the bias introduced by using a single Monte Carlo for both evaluating the $\alpha$s and the $\brcva$ price is negligible.

\section{A case study}\label{sec:casestudyrates}

In the following numerical examples we use as free correlation parameters:
\[
{\bar\rho}_C \;,\quad {\bar\rho}_I \;,\quad \rho_G \ .
\]
and we recover the other correlations from them. In particular, we consider the following cases:
\begin{itemize}
\item Varying ${\bar\rho}_C$, keeping fixed ${\bar\rho}_I=0$.
\item Varying both ${\bar\rho}_C$ and ${\bar\rho}_I$, keeping them equal, i.e. ${\bar\rho}_C={\bar\rho}_I$.
\item For each choice of ${\bar\rho}_C$ and ${\bar\rho}_I$, we consider $\rho_G \in \{-80\%,0\%,80\%\}$.
\end{itemize}

We consider payoffs depending on at-the-money forward interest-rate-swap (IRS) paying on the EUR market. These contracts reset a given number of years from trade date and start accruing two business days later. The IRS's fixed legs pay annually a 30e/360 strike rate, while the floating legs pay LIBOR twice per year.

In order to account for possible netting agreements, we consider three portfolios of swaps:
\begin{description}
\item[P1] A portfolio of 10 swaps, where all the swaps start at date $T_0$ and the $i$-th swap matures $i$ years after the starting date. The netting of the portfolio is equal to an amortizing swap with decreasing outstanding.
\item[P2] A portfolio of 10 swaps, where all the swaps mature in 10 years from date $T_0$, but they start at different dates, namely the $i$-th swap starts $i-1$ years from date $T_0$. The netting of the portfolio is equal to an amortizing swap with increasing outstanding.
\item[P3] A portfolio of 10 swaps, where all the swaps start at date $T_0$ and mature in 10 years. The netting of the portfolio is equal to a swap similar to the ones in the portfolio but with 10 times larger notional.
\end{description}

\begin{center}
\scalebox{0.7}{
Portfolio P1
\begin{tikzpicture}
\draw[->, color=gray, line width=3pt] (0,6) -- (12,6);
\draw[color=red, line width=4pt] (0,6-0.1) -- (0,6+0.1);
\node at (0, 6+0.3) {$t_0$};
\draw[->, color=blue, line width=2pt] (1.0,6) -- (1,9) node[right, text width=4em] {$10K$};
\node at (1, 6-0.3) {$t_1$};
\draw[->, color=blue, line width=2pt] (2.0,6) -- (2,8.7) node[right, text width=4em] {$9K$};
\node at (2, 6-0.3) {$t_2$};
\draw[->, color=blue, line width=2pt] (3.0,6) -- (3,8.4) node[right, text width=4em] {$8K$};
\node at (3, 6-0.3) {$t_3$};
\draw[->, color=blue, line width=2pt] (4.0,6) -- (4,8.1) node[right, text width=4em] {$7K$};
\node at (4, 6-0.3) {$t_4$};
\draw[->, color=blue, line width=2pt] (5.0,6) -- (5,7.8) node[right, text width=4em] {$6K$};
\node at (5, 6-0.3) {$t_5$};
\draw[->, color=blue, line width=2pt] (6,6) -- (6,7.5) node[right, text width=4em] {$5K$};
\node at (6, 6-0.3) {$t_6$};
\draw[->, color=blue, line width=2pt] (7,6) -- (7,7.2) node[right, text width=4em] {$4K$};
\node at (7, 6-0.3) {$t_7$};
\draw[->, color=blue, line width=2pt] (8,6) -- (8,6.9) node[right, text width=4em] {$3K$};
\node at (8, 6-0.3) {$t_8$};
\draw[->, color=blue, line width=2pt] (9,6) -- (9,6.6) node[right, text width=4em] {$2K$};
\node at (9, 6-0.3) {$t_9$};
\draw[->, color=blue, line width=2pt] (10,6) -- (10,6.3) node[right, text width=4em] {$1K$};
\node at (10, 6-0.3) {$t_{10}$};
\end{tikzpicture}
}
\scalebox{0.7}{
Portfolio P2
\begin{tikzpicture}
\draw[->, color=gray, line width=3pt] (0,6) -- (12,6);
\draw[color=red, line width=4pt] (0,6-0.1) -- (0,6+0.1);
\node at (0, 6+0.3) {$t_0$};
\draw[->, color=blue, line width=2pt] (1.0,6) -- (1,6.3) node[right, text width=4em] {$K$};
\node at (1, 6-0.3) {$t_1$};
\draw[->, color=blue, line width=2pt] (2.0,6) -- (2,6.6) node[right, text width=4em] {$2K$};
\node at (2, 6-0.3) {$t_2$};
\draw[->, color=blue, line width=2pt] (3.0,6) -- (3,6.9) node[right, text width=4em] {$3K$};
\node at (3, 6-0.3) {$t_3$};
\draw[->, color=blue, line width=2pt] (4.0,6) -- (4,7.2) node[right, text width=4em] {$4K$};
\node at (4, 6-0.3) {$t_4$};
\draw[->, color=blue, line width=2pt] (5.0,6) -- (5,7.5) node[right, text width=4em] {$5K$};
\node at (5, 6-0.3) {$t_5$};
\draw[->, color=blue, line width=2pt] (6,6) -- (6,7.8) node[right, text width=4em] {$6K$};
\node at (6, 6-0.3) {$t_6$};
\draw[->, color=blue, line width=2pt] (7,6) -- (7,8.1) node[right, text width=4em] {$7K$};
\node at (7, 6-0.3) {$t_7$};
\draw[->, color=blue, line width=2pt] (8,6) -- (8,8.4) node[right, text width=4em] {$8K$};
\node at (8, 6-0.3) {$t_8$};
\draw[->, color=blue, line width=2pt] (9,6) -- (9,8.7) node[right, text width=4em] {$9K$};
\node at (9, 6-0.3) {$t_9$};
\draw[->, color=blue, line width=2pt] (10,6) -- (10,9) node[right, text width=4em] {$10K$};
\node at (10, 6-0.3) {$t_{10}$};
\end{tikzpicture}
}
\scalebox{0.7}{
Portfolio P3
\begin{tikzpicture}
\draw[->, color=gray, line width=3pt] (0,6) -- (12,6);
\draw[color=red, line width=4pt] (0,6-0.1) -- (0,6+0.1);
\node at (0, 6+0.3) {$t_0$};
\draw[->, color=blue, line width=2pt] (1.0,6) -- (1,9) node[right, text width=4em] {$10K$};
\node at (1, 6-0.3) {$t_1$};
\draw[->, color=blue, line width=2pt] (2.0,6) -- (2,9) node[right, text width=4em] {$10K$};
\node at (2, 6-0.3) {$t_2$};
\draw[->, color=blue, line width=2pt] (3.0,6) -- (3,9) node[right, text width=4em] {$10K$};
\node at (3, 6-0.3) {$t_3$};
\draw[->, color=blue, line width=2pt] (4.0,6) -- (4,9) node[right, text width=4em] {$10K$};
\node at (4, 6-0.3) {$t_4$};
\draw[->, color=blue, line width=2pt] (5.0,6) -- (5,9) node[right, text width=4em] {$10K$};
\node at (5, 6-0.3) {$t_5$};
\draw[->, color=blue, line width=2pt] (6,6) -- (6,9) node[right, text width=4em] {$10K$};
\node at (6, 6-0.3) {$t_6$};
\draw[->, color=blue, line width=2pt] (7,6) -- (7,9) node[right, text width=4em] {$10K$};
\node at (7, 6-0.3) {$t_7$};
\draw[->, color=blue, line width=2pt] (8,6) -- (8,9) node[right, text width=4em] {$10K$};
\node at (8, 6-0.3) {$t_8$};
\draw[->, color=blue, line width=2pt] (9,6) -- (9,9) node[right, text width=4em] {$10K$};
\node at (9, 6-0.3) {$t_9$};
\draw[->, color=blue, line width=2pt] (10,6) -- (10,9) node[right, text width=4em] {$10K$};
\node at (10, 6-0.3) {$t_{10}$};
\end{tikzpicture}
}
\end{center}

We analyze the impact of correlations, interest-rate curve and credit spreads level and volatility scenarios on bilateral CVA calculations.
\clearpage

\subsection{Main findings}

In general our results confirm, both in the mid- and in the high-risk settings, the bilateral credit valuation adjustment to be relevant and structured. We in particular notice that the impact of correlations between investor's and counterparty's default risks is relevant. We also find a relevant impact of credit spread volatilities for the credit qualities of both names, and of correlation between defaults and interest rates, as was earlier found for unilateral CVA calculations in Brigo and Pallavicini (2007).

Also, in several scenarios the value of $\brcva$ may change sign according to the investor's and the counterparty's credit risk level and volatilities and depending on the correlation of these risks with the interest rates. This change of sign feature is a further convincing reason of the impact of dynamics on rigorous CVA valuation. The possible change of sign is also unique of the bilateral case, the unilateral adjustment having always the same sign.

We are going to detail our findings in a number of illustrative examples from our extensive set of results.

\subsection{Netted IRS portfolios: right- and wrong-way risk}

Table~\ref{tab:hmrisk} reports a first panel of results. It is the bilateral credit valuation adjustment for three different receiver IRS portfolios with ten years maturity, using the high-risk parameter set for the counterparty credit spread and the mid-risk parameter set for the investor credit spread. The two default times are assumed uncorrelated, $\rho_G=0$. This first set considers the CVA calculation for the three different portfolios for a number of possible behaviours of wrong-way correlations.

When $\bar{\rho}_I$ is kept to zero, we notice the same pattern in $\bar{\rho}_C$ we had seen in the unilateral case in Brigo and Pallavicini (2007). Increasing correlation $\bar{\rho}_C$ means that, ceteris paribus, higher interest rates will correspond to high credit spreads, putting the receiver swaptions embedded in the $\lgd_C$ term of the adjustment more out of the money. This will cause the $\lgd_C$ term of the adjustment to diminish in absolute value, so that the final value of the CVA will be larger for high correlation. This is clearly seen in the left panel of Table~\ref{tab:hmrisk}, where the CVA is seen to increase as $\bar{\rho}_C$ increases, given that the table columns increase.

When, in the right panel of Table~\ref{tab:hmrisk},  $\bar{\rho}_I$ is taken to follow $\bar{\rho}_C$, the behaviour is the same but more marked. This is reasonable: when $\bar{\rho}_C$ is large, also $\bar{\rho}_I$ is now large. This means that, ceteris paribus, higher interest rates will correspond to high credit spreads, putting the payer swaptions embedded in the $\lgd_I$ term of the adjustment more in the money, so that this term is larger. This makes the CVA increase further. Not surprisingly, the numbers in the right panel of Table~\ref{tab:hmrisk} corresponding to positive correlation (bottom part of the table) are all larger than the corresponding numbers in the left panel.

It is worth finally checking the impact of correlation on the CVA, comparing it with the typical $[1.2, 1.4]$ interval adjustment factor postulated by Basel II for the credit risk measurement correction due to wrong-way risk. Depending on whether we look at the deal from the Investor or Counterparty point of view, we find the following ratios between nonzero correlation CVA and zero correlation CVA. For example, we find
\[
382/148 \approx 2.58, \ \ 159 / 31 \approx 5.13
\]
which are both much larger than $1.4$. This means that mimicking the Basel II rules in the valuation space is not going to work, since the impact of correlations and volatilities is much more complex than what can be achieved with a simple multiplier.

Finally, we notice that depending on the correlations $\bar{\rho}_I, \bar{\rho}_C$ the CVA does change sign, and in particular for portfolios P1 and P3 the sign of the adjustment follows the sign of the correlations. P2 is an exception because of the more massive presence of cash flows in the future.

\begin{table}[htp]
\begin{center}
\begin{minipage}{0.48\textwidth}
\centering
{\small
\begin{tabular}{|cc|ccc|}\hline
$\boldsymbol {\bar\rho}_C$ & $\boldsymbol {\bar\rho}_I$ &   {\bf P1} &   {\bf P2} &   {\bf P3} \\\hline
{\bf -60\%} &  {\bf 0\%} &       -117(7)&       -382(12) &       -237(16) \\
{\bf -40\%} &  {\bf 0\%} &       -74(6) &       -297(11) &       -138(15) \\
{\bf -20\%} &  {\bf 0\%} &       -32(6) &       -210(10) &        -40(14) \\\hline
{\bf 0\%} &  {\bf 0\%} &          -1(5) &       -148(9) &         31(13) \\\hline
{\bf 20\%} &  {\bf 0\%} &         24(5) &        -96(9) &         87(12) \\
{\bf 40\%} &  {\bf 0\%} &         44(4) &        -50(8) &        131(11) \\
{\bf 60\%} &  {\bf 0\%} &         57(4) &        -22(7) &        159(11) \\\hline
\end{tabular}}
\end{minipage}
\begin{minipage}{0.48\textwidth}
\centering
{\small
\begin{tabular}{|cc|ccc|}\hline
$\boldsymbol {\bar\rho}_C$ & $\boldsymbol {\bar\rho}_I$ &   {\bf P1} &   {\bf P2} &   {\bf P3} \\\hline
{\bf -60\%} & {\bf -60\%} &       -150(6) &       -422(12) &       -319(15) \\
{\bf -40\%} & {\bf -40\%} &        -98(6) &       -329(11) &       -197(14) \\
{\bf -20\%} & {\bf -20\%} &        -46(5) &       -230(10) &        -74(13) \\\hline
{\bf 0\%} &  {\bf 0\%} &            -1(5) &       -148(9) &         31(13) \\\hline
{\bf 20\%} & {\bf 20\%} &         38(5) &        -77(9) &        121(12) \\
{\bf 40\%} & {\bf 40\%} &         75(5) &         -6(8) &        208(12) \\
{\bf 60\%} & {\bf 60\%} &        106(5) &         49(8) &        280(12) \\\hline
\end{tabular}}
\end{minipage}
\caption{Bilateral credit valuation adjustment for three different receiver IRS portfolios for a maturity of ten years, using high-risk parameter set for the counterparty and mid-risk parameter set for the investor with uncorrelated default times. Every IRS has unitary notional. Prices are in basis points.}\label{tab:hmrisk}
\end{center}
\end{table}

\subsection{Netted portfolios and credit spreads}

In Table~\ref{tab:p1leftp2right} we report our second example of relevant results. We analyze the  bilateral credit valuation adjustment for the two portfolios P1 and P2, again with uncorrelated default times, $\rho_G =0$.

Here too, depending on the correlations $\bar{\rho}_I, \bar{\rho}_C$, we see that the CVA may change sign. Also, we notice that two examples of wrong-way risk we would get, as ratios of high (positive or negative) correlation CVA over zero correlation CVA, are
\[
422/148 \approx 2.85, \ \ 315/16 \approx 19.7
\]
which are dramatically larger than $1.4$.

We also notice, in the table, as we move left to right along one row, that the CVA always grows. This is expected, since we are looking at the CVA adjustment to be added by the investor. This way the configuration where the counterparty has high spread risk and the investor medium spread risk will produce smaller CVA's with respect to the case where both investor and counterparty have high spread risk. This is because in the case where the investor has medium spread risk default times of the investor will tend to be later than in the case where the investor has high spread risk. Therefore, investor default probabilities will be larger in the latter case of high investor spread risk, and as a consequence the $\lgd_I$ term in the adjustment will be larger in the latter case. Since this term is positive in the adjustment to be added by the investor, this will produce a larger bilateral CVA.
\clearpage

\begin{table}[htp]
\begin{center}
\begin{minipage}{0.495\textwidth}
\centering
{\small
\begin{tabular}{|cc|ccc|}\hline
$\boldsymbol {\bar\rho}_C$ & $\boldsymbol {\bar\rho}_I$ &  {\bf H/M} &  {\bf H/H} &  {\bf M/H} \\\hline
{\bf -60\%} & {\bf -60\%} &       -150(6) &        -76(7) &         47(5) \\
{\bf -40\%} & {\bf -40\%} &        -98(6) &        -12(6) &         97(5) \\
{\bf -20\%} & {\bf -20\%} &        -46(5) &         48(6) &        135(5) \\\hline
{\bf 0\%} &  {\bf 0\%} &          -1(5) &        110(6) &        187(6) \\\hline
{\bf 20\%} & {\bf 20\%} &         38(5) &        173(6) &        241(6) \\
{\bf 40\%} & {\bf 40\%} &         75(5) &        239(7) &        297(6) \\
{\bf 60\%} & {\bf 60\%} &        106(5) &        304(7) &        361(7) \\
\hline
\end{tabular}}
\end{minipage}
\begin{minipage}{0.495\textwidth}
\centering
{\small
\begin{tabular}{|cc|ccc|}\hline
$\boldsymbol {\bar\rho}_C$ & $\boldsymbol {\bar\rho}_I$ &  {\bf H/M} &  {\bf H/H} &  {\bf M/H} \\\hline
{\bf -60\%} & {\bf -60\%} &       -422(12) &       -284(12) &        -40(9) \\
{\bf -40\%} & {\bf -40\%} &       -329(11) &       -179(12) &         36(9) \\
{\bf -20\%} & {\bf -20\%} &       -230(10) &        -77(10) &        102(9) \\\hline
{\bf 0\%} &  {\bf 0\%} &       -148(10) &         16(10) &        179(9) \\\hline
{\bf 20\%} & {\bf 20\%} &        -77(9) &        112(10) &        262(10) \\
{\bf 40\%} & {\bf 40\%} &         -6(9) &        218(10) &        351(10) \\
{\bf 60\%} & {\bf 60\%} &         49(9) &        315(11) &        450(11) \\\hline
\end{tabular}}
\end{minipage}
\caption{Bilateral credit valuation adjustment, by changing the parameter set, for a decreasing (P1, left panel) and an increasing (P2, right panel) IRS portfolio for a maturity of ten years, with uncorrelated default times. Every IRS has unitary notional. Prices are in basis points.}\label{tab:p1leftp2right}
\end{center}
\end{table}

\subsection{Netted portfolios and default correlation}

In Table~\ref{tab:netdefcorr} we focus on the impact of $\rho_G$ on the adjustment. This is rich in structure and complex. Indeed, we see for example that, depending on the particular values of $\bar{\rho}_I$ and $\bar{\rho}_C$, an increase of $\rho_G$ can imply either an increase or a decrease of the adjustment. Even when staying with just negative $\bar{\rho}_I$ and $\bar{\rho}_C$'s this happens, as one can see by comparing the first and third row in the right panel of the table.

\begin{table}[htp]
\begin{center}
\begin{minipage}{0.49\textwidth}
\centering
{\small
\begin{tabular}{|cc|ccc|}\hline
$\boldsymbol {\bar\rho}_C$ & $\boldsymbol {\bar\rho}_I$ & {\bf -80\%} &  {\bf 0\%} & {\bf 80\%} \\\hline
{\bf -60\%} & {\bf -60\%} &       -150(7) &       -150(6) &       -169(6) \\
{\bf -40\%} & {\bf -40\%} &        -91(6) &        -98(6) &       -122(6) \\
{\bf -20\%} & {\bf -20\%} &        -33(6) &        -46(5) &        -72(5) \\\hline
{\bf 0\%} &  {\bf 0\%} &          18(6) &           -1(5) &        -34(5) \\\hline
{\bf 20\%} & {\bf 20\%} &         61(5) &           38(5) &         -3(4) \\
{\bf 40\%} & {\bf 40\%} &        102(5) &           75(5) &         29(4) \\
{\bf 60\%} & {\bf 60\%} &        140(5) &          106(5) &         53(4) \\\hline
\end{tabular}}
\end{minipage}
\begin{minipage}{0.49\textwidth}
\centering
{\small
\begin{tabular}{|cc|ccc|}\hline
$\boldsymbol {\bar\rho}_C$ & $\boldsymbol {\bar\rho}_I$ & {\bf -80\%} &  {\bf 0\%} & {\bf 80\%} \\\hline
{\bf -60\%} & {\bf -60\%} &         32(5) &         47(5) &         61(5) \\
{\bf -40\%} & {\bf -40\%} &         86(5) &         97(5) &        103(5) \\
{\bf -20\%} & {\bf -20\%} &        146(6) &        135(5) &        137(5) \\\hline
{\bf 0\%} &  {\bf 0\%} &        194(6) &        187(6) &        183(5) \\\hline
{\bf 20\%} & {\bf 20\%} &        256(6) &        241(6) &        232(6) \\
{\bf 40\%} & {\bf 40\%} &        320(7) &        297(6) &        287(6) \\
{\bf 60\%} & {\bf 60\%} &        384(7) &        361(7) &        344(7) \\\hline
\end{tabular}}
\end{minipage}
\caption{Bilateral credit valuation adjustment, by changing the Gaussian copula parameter $\rho_G$ for a decreasing IRS portfolio (P1) for a maturity of ten years, using high-risk parameter set for the counterparty and mid-risk parameter set for the investor (left panel), and inverted settings (right panel). Every IRS has unitary notional. Prices are in basis points.}\label{tab:netdefcorr}
\end{center}
\end{table}


\subsection{Netted portfolios and volatility of credit spreads}

Table \ref{tab:netSpreadsVol} illustrates the impact of the Counterparty's credit spreads volatility on the adjustment. We use high-risk credit spreads for the counterparty and mid-risk credit spreads and parameter set for the investor. Every time we change the main volatility parameter $\nu^C$ in the counterparty credit spread model, we apply a shift $\psi^C(t,\beta^C)$ to fit the credit spread curve of the high-risk scenario, so that the credit spread model for the counterparty fits the same initial high-risk CDS spread curve even if altering the credit spread volatility.

Our results highlight once again the importance of credit spread volatilities, often neglected in the literature. The adjustment can even double by a change in volatility. Assuming zero volatility is a quite strong tacit assumption that is certainly not granted by CDS volatilities, either implied or historical (see for example Brigo (2005)).

\begin{table}[h!]
\begin{center}
\begin{minipage}{0.49\textwidth}
\centering
{\small
\begin{tabular}{|cc|ccc|}\hline
$\boldsymbol {\bar\rho}_C$ & $\boldsymbol {\bar\rho}_I$ & {\bf 10\%} &  {\bf 30\%} & {\bf 50\%} \\\hline
{\bf -60\%} & {\bf -60\%} &       -70(5) &      -128(6) &      -150(6) \\
{\bf -40\%} & {\bf -40\%} &       -48(5) &       -84(6) &       -98(6) \\
{\bf -20\%} & {\bf -20\%} &       -25(5) &       -43(5) &       -46(5) \\\hline
{\bf 0\%} &  {\bf 0\%} &           -2(5) &        -2(5) &        -1(5) \\\hline
{\bf 20\%} & {\bf 20\%} &          23(5) &        35(5) &        38(5) \\
{\bf 40\%} & {\bf 40\%} &          51(5) &        71(5) &        75(5) \\
{\bf 60\%} & {\bf 60\%} &          76(5) &       106(5) &       106(5) \\\hline
\end{tabular}}
\end{minipage}
\begin{minipage}{0.49\textwidth}
\centering
{\small
\begin{tabular}{|cc|ccc|}\hline
$\boldsymbol {\bar\rho}_C$ & $\boldsymbol {\bar\rho}_I$ & {\bf 10\%} &  {\bf 30\%} & {\bf 50\%} \\\hline
{\bf -60\%} & {\bf -60\%} &      -276(10) &        -396(11) &       -422(12) \\
{\bf -40\%} & {\bf -40\%} &      -238(10) &        -313(11) &       -329(11) \\
{\bf -20\%} & {\bf -20\%} &      -194(10) &        -235(10) &       -230(10) \\\hline
{\bf 0\%} &  {\bf 0\%} &         -151(10) &        -155(10) &       -148(10) \\\hline
{\bf 20\%} & {\bf 20\%} &        -109(9) &          -82(9) &         -77(9) \\
{\bf 40\%} & {\bf 40\%} &         -58(9) &          -12(9) &         	-6(9) \\
{\bf 60\%} & {\bf 60\%} &          -8(9) &           58(8) &          49(9) \\\hline
\end{tabular}}
\end{minipage}
\caption{Bilateral credit valuation adjustment, by changing the volatility $\nu^C$ of Counterparty's credit spreads for a decreasing IRS portfolio (P1, left panel), for a maturity of ten years, and increasing portfolio (P2, right panel). We use high-risk credit spreads for the counterparty and mid-risk credit spreads and parameter set for the investor. Every IRS has unitary notional. Prices are in basis points.}\label{tab:netSpreadsVol}
\end{center}
\end{table}

\subsection{Netted portfolios and forward rates}

Table~\ref{tab:netfwdrates} illustrates the impact of the shape of the initial interest-rate curve across maturity on the adjustment. We compare three possible shapes: increasing, flat and decreasing curves. We run this for the flat portfolio P3, but results will be possibly more dramatic for P2. As we see from the numbers in the table, the adjustment is quite sensitive to the shape of the initial curve.

\begin{table}[htp]
\begin{center}
\begin{minipage}{0.495\textwidth}
\centering
{
\footnotesize
\begin{tabular}{|cc|ccc|}\hline
$\boldsymbol {\bar\rho}_C$ & $\boldsymbol {\bar\rho}_I$ & {\bf Incr.} &  {\bf Flat} & {\bf Decr.} \\\hline
{\bf -60\%} & {\bf -60\%} &     -319(15) &        -777(18) &       -1193(21) \\
{\bf -40\%} & {\bf -40\%} &     -197(14) &        -630(17) &       -1032(20) \\
{\bf -20\%} & {\bf -20\%} &      -74(13) &        -472(15) &        -852(18) \\\hline
{\bf 0\%} &  {\bf 0\%} &          31(13) &        -344(14) &        -709(17) \\\hline
{\bf 20\%} & {\bf 20\%} &        121(12) &        -228(13) &        -571(15) \\
{\bf 40\%} & {\bf 40\%} &        208(12) &        -115(12) &        -436(14) \\
{\bf 60\%} & {\bf 60\%} &        280(12) &         -25(12) &        -328(13) \\\hline
\end{tabular}}
\end{minipage}
\begin{minipage}{0.495\textwidth}
\centering
{
\footnotesize
\begin{tabular}{|cc|ccc|}\hline
$\boldsymbol {\bar\rho}_C$ & $\boldsymbol {\bar\rho}_I$ & {\bf Incr.} &  {\bf Flat} & {\bf Decr.} \\\hline
{\bf -60\%} & {\bf -60\%} &     169(13) &        -140(13) &       -410(15) \\
{\bf -40\%} & {\bf -40\%} &     288(13) &        -46(13) &       -325(14) \\
{\bf -20\%} & {\bf -20\%} &     384(13) &         40(12) &        -247(13) \\\hline
{\bf 0\%} &  {\bf 0\%} &        507(14) &         142(13) &        -159(13) \\\hline
{\bf 20\%} & {\bf 20\%} &       637(15) &         251(13) &        -66(13) \\
{\bf 40\%} & {\bf 40\%} &       773(16) &         374(14) &        42(14) \\
{\bf 60\%} & {\bf 60\%} &       925(17) &         511(15) &        166(14) \\\hline
\end{tabular}}
\end{minipage}
\caption{Bilateral credit valuation adjustment, by changing the yield curve (increasing, flat at $3\%$ and decreasing curves) for a flat IRS portfolio (P3) for a maturity of ten years, using high-risk parameter set for the counterparty and mid-risk parameter set for the investor (left panel), and inverted settings (right panel). We also assume uncorrelated default times. Every IRS has unitary notional. Prices are in basis points.}\label{tab:netfwdrates}
\end{center}
\end{table}

\subsection{Exotics}

In Table~\ref{tab:exoticsbrcva} we show the adjustment for exotic options on interest-rates. In particular we focus on options whose payoff may change sign depending on future fixing of quoted rates. The calculation of one-sided CVA for exotic interest-rate options is covered in Brigo and Pallavicini (2007). Here we address the bilateral case.

For instance, we consider IRS portfolio P3, and we add an auto-callable feature triggered by the Libor rate, namely we exit form the IRS contract when on a fix-leg payment date the fixing of the Libor rate is greater then a strike level $A$. We can appreciate that also for this product the correlations have quite an impact on the value of the adjustment.

\begin{table}[htp]
\begin{center}
{\small
\begin{tabular}{|cc|ccc|}\hline
$\boldsymbol {\bar\rho}_C$ & $\boldsymbol {\bar\rho}_I$ & {\bf -99\%} &  {\bf 0\%} & {\bf 99\%} \\\hline
{\bf -70\%} & {\bf -70\%} &        -71 &        -64 &        -55 \\
{\bf 0\%} &  {\bf 0\%} &        -47 &        -43 &        -34 \\
{\bf 70\%} & {\bf 70\%} &        -28 &        -26 &        -20 \\\hline
\end{tabular}}
\caption{Bilateral credit valuation adjustment, by changing the Gaussian copula parameter $\rho_G$, for an auto-callable IRS portfolio (P3) for a maturity of ten years, using high-risk parameter set for the counterparty and mid-risk parameter set for the investor. The contract has unitary notional. Prices are in basis points. Intrinsic price is $608$, with an auto-callable strike level of $A=3\%$.}\label{tab:exoticsbrcva}
\end{center}
\end{table}

\section{Further discussion and conclusions}\label{sec:conclbilateralrates}

In general our results confirm the bilateral credit valuation adjustment to be quite sensitive to finely tuned dynamics parameters such as volatilities and correlations, similarly to what was found in Brigo and Capponi (2008) for the CDS market. The impact of the parameters is both relevant and structured.

We in particular noticed the impact of correlations between investor's and counterparty's default risks, of credit spread volatilities for the credit qualities of both names, of credit spread levels  and of correlations between defaults and interest rates. Variations in these parameters can produce an excursion in the adjustment of several multiples or even have the adjustment changing in sign.

In particular, there seems to be no single multiple that can provide the adjustment for high correlations starting from the adjustment with zero correlations. Hence the need to include such correlations in the modeling apparatus in a rigorous way. We proposed a possible modeling choice for addressing this, with a two-factor Gaussian model (G2++) for interest rates and shifted square root processes with possible jumps (JCIR++) for the credit spreads of investor and counterparty. Defaults of the two names are linked by a Gaussian copula function.

We detailed our findings in a number of illustrative examples from an extensive set of results.

\newpage

\appendix

\section{Interest-rate market data}\label{app:rates}

We report the yield curve term structure and swaption volatilities used to calibrate the interest-rate and credit-spread dynamics in Sections~\ref{subsec:interestrates} and~\ref{subsec:creditspreads}.

\begin{table}[htp]
\begin{center}
{\small
\begin{tabular}{|cc|cc|cc|}
\hline
{\bf Date} & {\bf Rate} & {\bf Date} & {\bf Rate} & {\bf Date} & {\bf Rate} \\\hline
 27-May-09 &     1.15\% &  28-Dec-09 &     1.49\% &  29-May-17 &     3.40\% \\
 28-May-09 &     1.02\% &  28-Jan-10 &     1.53\% &  28-May-18 &     3.54\% \\
 29-May-09 &     0.98\% &  26-Feb-10 &     1.56\% &  28-May-19 &     3.66\% \\
 04-Jun-09 &     0.93\% &  29-Mar-10 &     1.59\% &  28-May-21 &     3.87\% \\
 11-Jun-09 &     0.92\% &  28-Apr-10 &     1.61\% &  28-May-24 &     4.09\% \\
 18-Jun-09 &     0.91\% &  28-May-10 &     1.63\% &  28-May-29 &     4.19\% \\
 29-Jun-09 &     0.91\% &  30-May-11 &     1.72\% &  29-May-34 &     4.07\% \\
 28-Jul-09 &     1.05\% &  28-May-12 &     2.13\% &  30-May-39 &     3.92\% \\
 28-Aug-09 &     1.26\% &  28-May-13 &     2.48\% &  & \\
 28-Sep-09 &     1.34\% &  28-May-14 &     2.78\% &  & \\
 28-Oct-09 &     1.41\% &  28-May-15 &     3.02\% &  & \\
 30-Nov-09 &     1.46\% &  30-May-16 &     3.23\% &  & \\\hline
\end{tabular}}
\caption{\label{disc}
EUR zero-coupon continuously-compounded spot rates (ACT/360) observed on May, 26 2009.
}
\end{center}
\end{table}

\begin{table}[htp]
\begin{center}
{\small
\begin{tabular}{|c|cccccccccc|}\hline
{\bf $\boldsymbol t\downarrow$ / $\boldsymbol b\rightarrow$}   & {\bf 1y} & {\bf 2y} & {\bf 3y} & {\bf 4y} & {\bf 5y} & {\bf 6y} & {\bf 7y} & {\bf 8y} & {\bf 9y} & {\bf 10y} \\\hline
{\bf 1y} &    42.8\% &    34.3\% &    31.0\% &    28.8\% &    27.7\% &    26.9\% &    26.5\% &    26.3\% &    26.2\% &    26.2\% \\
{\bf 2y} &    28.7\% &    25.6\% &    24.1\% &    23.1\% &    22.4\% &    22.3\% &    22.2\% &    22.3\% &    22.4\% &    22.4\% \\
{\bf 3y} &    23.5\% &    21.1\% &    20.4\% &    20.0\% &    19.7\% &    19.7\% &    19.7\% &    19.8\% &    19.9\% &    20.1\% \\
{\bf 4y} &    19.9\% &    18.5\% &    18.2\% &    18.1\% &    18.0\% &    18.1\% &    18.1\% &    18.2\% &    18.2\% &    18.4\% \\
{\bf 5y} &    17.6\% &    16.8\% &    16.9\% &    16.9\% &    17.0\% &    16.9\% &    17.0\% &    17.0\% &    17.0\% &    17.1\% \\
{\bf 7y} &    15.4\% &    15.3\% &    15.3\% &    15.3\% &    15.3\% &    15.3\% &    15.3\% &    15.4\% &    15.5\% &    15.6\% \\
{\bf 10y} &   14.2\% &    14.2\% &    14.2\% &    14.3\% &    14.4\% &    14.5\% &    14.6\% &    14.7\% &    14.8\% &    15.0\% \\\hline
\end{tabular}}
\caption{\label{swapvol}
Market at-the-money swaption volatilities observed on May, 26 2009. Each column contains volatilities of swaptions of a given tenor $b$ for different expiries $t$.
}
\end{center}
\end{table}

\section{CDS Terms Structures and Implied Volatilities}\label{app:credit}

We report the CDS term structures and implied volatilities associated with the model parameter used for the credit-spread dynamics in Section~\ref{subsec:creditspreads}.


\begin{table}[htp]
\begin{center}
\begin{tabular}{|c|cccccccccc|}\hline
{$\boldsymbol T$} &    {\bf 1y} &    {\bf 2y} & {\bf 3y} & {\bf 4y} & {\bf 5y} & {\bf 6y} & {\bf 7y} & {\bf 8y} & {\bf 9y} &  {\bf 10y} \\\hline
{\bf CDS Spread} &     92 &        104 &        112 &        117 &        120 &        122 &        124 &        125 &        126 &        127 \\\hline
\end{tabular}
\caption{Mid risk initial CDS term structure}\label{tab:cdstermmid}
\end{center}
\end{table}

\begin{table}[htp]
\begin{center}
{\small
\begin{tabular}{|c|ccccccccc|}\hline
{\bf $\boldsymbol t\downarrow$ / $\boldsymbol T\rightarrow$} &   {\bf 2y} &   {\bf 3y} &   {\bf 4y} &   {\bf 5y} &   {\bf 6y} &   {\bf 7y} &   {\bf 8y} &   {\bf 9y} &  {\bf 10y} \\\hline
  {\bf 1y} &       52\% &       36\% &       27\% &       21\% &       17\% &       15\% &       13\% &       12\% &       11\% \\
  {\bf 2y} &            &       39\% &       28\% &       21\% &       17\% &       14\% &       12\% &       11\% &       10\% \\
  {\bf 3y} &            &            &       33\% &       24\% &       18\% &       15\% &       12\% &       11\% &        9\% \\
  {\bf 4y} &            &            &            &       29\% &       21\% &       16\% &       13\% &       11\% &        9\% \\
  {\bf 5y} &            &            &            &            &       26\% &       19\% &       15\% &       12\% &       10\% \\
  {\bf 6y} &            &            &            &            &            &       24\% &       17\% &       13\% &       11\% \\
  {\bf 7y} &            &            &            &            &            &            &       23\% &       16\% &       12\% \\
  {\bf 8y} &            &            &            &            &            &            &            &       21\% &       15\% \\
  {\bf 9y} &            &            &            &            &            &            &            &            &       19\% \\\hline
\end{tabular}}
\caption{Mid risk CDS implied volatility associated to the parameters in Table~\ref{tab:yparammid1}. Each column contains volatilities of CDS options of a given maturity $T$ for different expiries $t$.}\label{tab:cdsvolsmid}
\end{center}
\end{table}


\begin{table}[htp]
\begin{center}
\begin{tabular}{|c|cccccccccc|}\hline
{$\boldsymbol T$} &  {\bf 1y} & {\bf 2y} & {\bf 3y} & {\bf 4y} & {\bf 5y} & {\bf 6y} & {\bf 7y} & {\bf 8y} & {\bf 9y} &  {\bf 10y} \\\hline
{\bf CDS Spread}  &   234     &      244 &      248 &      250 &      252 &      252 &      254 &      253 &      254 &        254 \\\hline
\end{tabular}
\caption{High risk initial CDS term structure}\label{tab:cdstermhigh}
\end{center}
\end{table}

\begin{table}[htp]
\begin{center}
{\small
\begin{tabular}{|c|ccccccccc|}\hline
{\bf $\boldsymbol t\downarrow$ / $\boldsymbol T\rightarrow$} &   {\bf 2y} &   {\bf 3y} &   {\bf 4y} &   {\bf 5y} &   {\bf 6y} &   {\bf 7y} &   {\bf 8y} &   {\bf 9y} &  {\bf 10y} \\\hline
  {\bf 1y} &       96\% &       69\% &       53\% &       43\% &       36\% &       31\% &       28\% &       26\% &       24\% \\
  {\bf 2y} &            &       71\% &       52\% &       40\% &       32\% &       27\% &       24\% &       21\% &       19\% \\
  {\bf 3y} &            &            &       59\% &       43\% &       33\% &       26\% &       22\% &       20\% &       18\% \\
  {\bf 4y} &            &            &            &       51\% &       37\% &       28\% &       23\% &       20\% &       17\% \\
  {\bf 5y} &            &            &            &            &       45\% &       33\% &       26\% &       21\% &       18\% \\
  {\bf 6y} &            &            &            &            &            &       40\% &       30\% &       24\% &       19\% \\
  {\bf 7y} &            &            &            &            &            &            &       40\% &       29\% &       22\% \\
  {\bf 8y} &            &            &            &            &            &            &            &       36\% &       26\% \\
  {\bf 9y} &            &            &            &            &            &            &            &            &       34\% \\\hline
\end{tabular}}
\caption{High risk CDS implied volatility associated to the parameters in Table~\ref{tab:yparamhigh1}. Each column contains volatilities of CDS options of a given maturity $T$ for different expiries $t$.}\label{tab:cdsvolshigh}
\end{center}
\end{table}


\newpage

\begin{thebibliography}{99}


\bibitem{assefa}
Assefa, S., Bielecki, T., Cr\'epey, S., and Jeanblanc, M. (2009).
CVA computation for counterparty risk assesment in credit portfolio. Preprint.

\bibitem{beumee} Beumee, J.,  Brigo, D.,  Schiemert, D., and Stoyle, G. (2009).
Charting a Course Through the CDS Big Bang. Fitch Solutions research report.

\bibitem{BieleckiJeanblanc}
Bielecki, T., Jeanblanc, M., and Rutkowski M. (2008).
Hedging of Credit Default Swaptions in a Hazard Process Model.
Available at \url{www.defaultrisk.com}.

\bibitem{patras}
Blanchet-Scalliet, C., and Patras, F. (2008).
Counterparty Risk Valuation for CDS.
Available at \url{www.defaultrisk.com}.

\bibitem{BrigoCDSMM}
Brigo, D. (2005).
Market Models for CDS Options and Callable Floaters.
{\em Risk}, January issue.

\bibitem{BrigoBudapest}
Brigo, D. (2008).
Counterparty Risk valuation with Stochastic Dynamical Models: Impact of Volatilities and Correlations.
Talk at fifth World Business Strategies Fixed Income Conference, Budapest, 26 September.

\bibitem{BrigoCapponi}
Brigo, D., and Capponi, A. (2008).
Bilateral counterparty risk valuation with stochastic dynamical models and application to Credit Default Swaps. Available at \url{ssrn.com} or at \url{arxiv.org}.

\bibitem {BrigoAlf05}
Brigo, D., and Alfonsi A. (2005).
Credit Default Swap Calibration and Derivatives Pricing with the SSRD Stochastic Intensity Model,
{\em Finance and Stochastic}, {\bf 9}, 29.

\bibitem{BrigoChourBakkar}
Brigo, D., and Bakkar I. (2009).
Accurate counterparty risk valuation for energy-commodities swaps.
{\em Energy Risk}, March issue.

\bibitem{BrigoBachir08}
Brigo, D., and El-Bachir, N. (2009).
An exact formula for default swaptions' pricing in the SSRJD stochastic intensity model.
{\em Mathematical Finance} in press.

\bibitem{Brigo08}
Brigo, D., and Chourdakis, K. (2008).
Counterparty Risk for Credit Default Swaps: Impact of spread volatility and default correlation.
{\em International Journal of Theoretical and Applied Finance} in press.

\bibitem{BrigoMas}
Brigo, D., and Masetti, M. (2005).
Risk Neutral Pricing of Counterparty Risk.
In Counterparty Credit Risk Modelling: Risk Management, Pricing and Regulation,
{\em Risk Books}, Pykhtin, M. editor, London.

\bibitem {Brigo06}
Brigo, D., and Mercurio, F. (2006).
Interest Rate Models: Theory and Practice -- with Smile, Inflation and Credit.
Second Edition, Springer Verlag, 2006.

\bibitem{BrigoPall}
Brigo, D., and Pallavicini A. (2007).
Counterparty Risk under Correlation between Default and Interest Rates.
In Numerical Methods for Finance,
{\em Chapman Hall}, Miller, J., Edelman, D., and Appleby, J. editors.


\bibitem{brigotarenghi} Brigo D., Tarenghi M. (2004), Credit Default Swap Calibration and Equity Swap
Valuation under Counterparty risk with a Tractable Structural Model, Working
Paper (Available at www.ssrn.com and www.defaultrisk.com).


\bibitem{Collin2002}
Collin-Dufresne P., Goldstein R., and Hugonnier J. (2004).
A general formula for pricing defaultable securities.
{\em Econometrica}, {\bf 72}, 1377.

\bibitem {Cox85}
Cox, J., Ingersoll, J., and Ross, S. (1985).
A theory of the term structure of interest rates.
{\em Econometrica}, {\bf 53}, 385.

\bibitem {Chou05}
Chourdakis, K. (2005).
Option pricing using the fractional FFT.
{\em Journal of Computational Finance}, {\bf 8}, 1.

\bibitem{crepey}
Cr\'epey, S., Jeanblanc, M., and B. Zargari (2009).
CDS with Counterparty Risk in a Markov Chain Copula Model with Joint Defaults.
Available at \url{www.defaultrisk.com}.

\bibitem{deprisco}
De Prisco, B., and Rosen, D. (2005).
Modelling Stochastic Counterparty Credit Exposures for Derivatives Portfolios.
In Counterparty Credit Risk Modelling: Risk Management, Pricing and Regulation,
{\em Risk Books}, Pykhtin, M. editor, London.

\bibitem{Duffie}
Duffie, D., and Lando, D. (2001).
Term structure of credit spreads with incomplete accounting information.
{\em Econometrica}, {\bf 63}, 633.

\bibitem{hille}
Hille, C.T., J. Ring and H. Shimanmoto (2005).
Modelling Counterparty Credit Exposure for Credit Default Swaps.
In Counterparty Credit Risk Modelling: Risk Management, Pricing and Regulation,
{\em Risk Books}, Pykhtin, M. editor, London.

\bibitem{Hull2000}
Hull, J., and White, A. (2000).
Valuing credit default swaps: Modelling default correlations.
Working paper, University of Toronto, 2000.


\bibitem{Leung}
Leung, S.Y., and Kwok, Y. K. (2005).
Credit Default Swap Valuation with Counterparty Risk.
{\em The Kyoto Economic Review}, {\bf 74}, 25.

\bibitem{lipton}
Lipton, A., and Sepp, A. (2009).
Credit value adjustment for credit default swaps via the structural default model.
{\em Journal of Credit Risk}, {\bf 5}, 123.

\bibitem{Picoult}
Picoult, E. (2005).
Calculating and Hedging Exposure, Credit Value Adjustment and Economic Capital for Counterparty Credit Risk
In Counterparty Credit Risk Modelling: Risk Management, Pricing and Regulation,
{\em Risk Books}, Pykhtin, M. editor, London.

\bibitem{Pyk}
Pykhtin, M. (2005).
Editor of Counterparty Credit Risk Modelling: Risk Management, Pricing and Regulation,
{\em Risk Books}, London.

\bibitem{Sorensen}
Sorensen, E. H. and Thierry F. Bollier (1994).
Pricing Swap Default Risk,
{\em Financial Analysts Journal}, {\bf 50}, 23.

\bibitem{Yi}
Yi, C. (2009).
Dangerous Knowledge: Credit Value Adjustment with Credit Triggers.
Bank of Montreal research paper.

\bibitem{Walker}
Walker, M. (2005).
Credit Default Swaps with Counterparty Risk: A Calibrated Markov Model.
Available at \url{www.physics.utoronto.ca/$\sim$qocmp/CDScptyNew.pdf}.

\end{thebibliography}
\end{document}